# Stoic Conceptual Modeling Applied to Business Process Modeling Notation (BPMN)

Sabah Al-Fedaghi*

*Computer Engineering Department*
*Kuwait University*
*Kuwait*

salfedaghi@yahoo.com, sabah.alfedaghi@ku.edu.kw

*Abstract* – **Basic abstraction principles are reached through ontology, which was traditionally conceived as a depiction of the world itself. Ontology is also described using conceptual modeling (CM) that defines fundamental concepts of reality. CM is one of the central activities in computer science, especially as it is mainly used in software engineering as an intermediate artifact for system construction. To achieve such a goal, we propose Stoic CM (SCM) as a description of what a system must do functionally with minimal ambiguity. As a case study, we apply SCM to investigate the ontology of BPMN (business process modeling notation). Such an undertaking would demonstrate SCM notions and simultaneously may offer a viable ontological foundation for BPMN. SCM defines the *being* of things and actions in reality based on Stoic notions of existence and subsistence. It has two levels of specification: (1) a subsistence static model where things and actions *subsist* and (2) an existence dynamic model where things and actions *exist* in time. From the Stoic ontological point of view, while a thing *existing* has a clear denotation, *subsistence* indicates the thing is "being there," but it is inactive (does not participate in an event). We apply SCM to BPMN processes that involve buying a new car with many notions, such as activity, task, event, and message. The result indicates that SCM produces a tighter representation of reality, thus providing the necessary description of the part in the application world to be used as requirements for developing the software system.**

## I. INTRODUCTION

Anthropologists think that abstraction of reality is the most important feature that gave *homo sapiens* a competitive edge over less developed human races. Basic abstraction principles are reached through ontology, which traditionally conceived is a description of the world itself. Ever since the Greek philosopher Aristotle, ontology has served as a basis for human theories and the construction of models [1]. Ontology is also used in conceptual modeling (CM) to define fundamental concepts of reality.

CM is one of the central activities in computer science, especially as it is mainly used in software engineering as an intermediate artifact for system construction [2]. In this paper, we propose Stoic CM (SCM) as a description of what a system must do functionally with minimal ambiguity [3].

---------------------------------------------------

We claim that Stoic ontology can provide a foundation for CM and as a way to evaluate the ontological soundness of a CM language and its corresponding concepts. Ontology refers to things whose *existence* is acknowledged by a system [4][5]. As a case study, we investigate applying SCM to business process modeling notation (BPMN). Such a venture would demonstrate the SCM notions and may simultaneously provide a possible ontological foundation for BPMN.

### A. Business Process Modeling Notation

The complexity of contemporary business information systems has motivated interest in studies that focus on CM as an aid to facilitate the comprehension of certain domain facts relating to such systems, which would contribute to better design decisions and eventually a better system [6]. Understanding the CM of business domains is a key skill for practitioners tasked with system analysis and design [6].

BPMN has been utilized to create 1) descriptive business processes models that can be communicated and analyzed and 2) technical view targets for technical developers who need detailed specifications on the models to make them executable [7]. BPMN enhances communication between business analysts, technical developers, and business people [8].

Theorists have proposed that BPMN can be used as a base for CM for its simplicity and expressiveness. Moreover, it is claimed that BPMN presents a well-defined semantic structure and provides an easy working platform [9]. According to [10], BPMN is characterized by its ability "to describe and reflect the real world of information systems better. This specification is comprehensive and partially conflicting." Therefore, "several researchers present an ontology that provides a formal definition of BPMN and can be used as a knowledge base" [10].

The focus of the BPMN standard is on providing an intuitive graphical language, rather than formal semantics specifications. This emphasis results in semantic ambiguities regarding the interpretation of its modeling constructs [11]. There is no guideline for using BPMN in ontology-based systems, and different communities have been working on BPMN-like ontologies for disparate application purposes [11]. The following example motivates further study of BPMN ontology.

### B. Sample Problem

Amongst main BPMN modeling notions is the notion of an event. In BPMN, events affect the flow of the model: "throw events cause something to happen; catch events are caused to happen. Moreover, depending on their position in the process flow, they are: start events, end events, or intermediate events" [11]. A BPMN event that references a named message is known as a *message event*. A message represents the content of communication between two participants [12].

Reference [11] scrutinized the BPMN diagram shown in Fig. 1. Messages are exchanged between the process pools by using tasks of the type, *send message*. BPMN includes *message* as a throw event, which is used to model the sending of a message as well. Since both *message task* (activity) and *message throw* (event) model the same sending of a message, there are differences in meaning. In general, an *event* conveys the idea that when something happens…, whereas an *activity* places more focus on the idea that something needs to be done. Events represent changes in the domain being modeled, whereas activities refer to a participant's commitment towards the fulfillment of a specific goal. An event maps to a time point on a timeline, whereas a task (or an activity) maps to a time interval. This relationship between BPMN constructs and the temporal line suggests that events are instantaneous while activities last in time. "BPMN does not commit to a theory of time points or intervals; every reference to time beyond atomicity remains vague within its modeling framework" [11].

### C. Paper Structure

This paper has two aims: 1) exploring the ontological features of SCM as a new modeling language and 2) applying SCM in a new application area, the BPMN. Accordingly, the paper is structured as follows. The next section gives a description of SCM that includes materials that have appeared in previous publications to provide a self-contained manuscript. Additionally, the paper includes a new contribution to SCM, which is an ongoing process. Section 3 gives samples of SCM modeling. Section 4 includes a BPMN case study modeled using SCM notions.

## II. SCM AND THINGING MACHINE MODEL

SCM utilizes thinging machine (TM) modeling [13] based on Stoic ontology [14] and Lupascian logic [15]. In TM modeling, a *thing* is a Heideggerian notion that indicates *something* that becomes itself and announces its existence or names the entering of the thing into the world (using Stoic ontology terms, see e.g., [16]). Such things are conceptualized as thimacs (*thi*ngs/*ma*chines). A thimac can be described similar to a Romer's system: "any part of reality, which can, at least in principle, be separated from the rest of the world and can be made an object of investigation" [17].

Additionally, a thimac (Fig. 2) has a dual mode of being: the machine side and the thing side. The machine has the five *potential* actions: create, process, release, transfer, and receive. The sense of '"machinery" originated in the TM five actions indicating that everything that creates, processes, and moves (release-transfer-receive) other things is a machine. Simultaneously, what a machine creates, processes (changes), releases, transfers, and/or receives is a thing. In this modeling view, the "world" is a totality of thimacs.

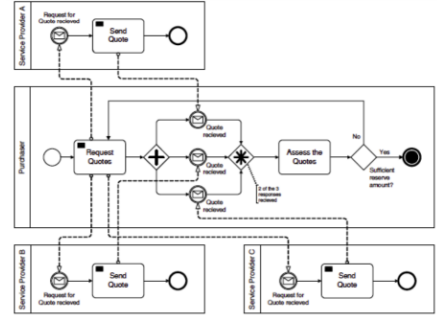

**Fig. 1 BPMN process diagram (From [12])**

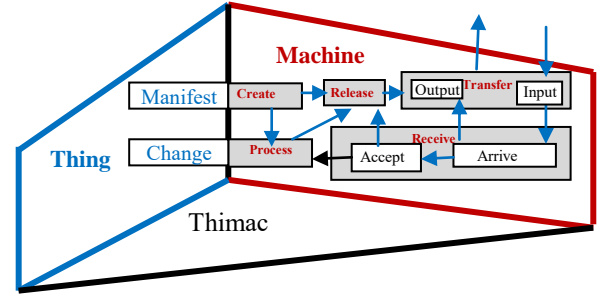

**Fig. 2. The thimac as a thing and machine**

TM modeling has two levels of specification:

*1) Static Model:* represents static things and static actions. From the SCM point of view, a thing's "being" at this level is a certain state of being, *subsistence* or a potential for "becoming," i.e., "it is there," inert, passive, waiting to exist when it couples with time. Becoming refers to transferring to the dynamic level to trigger creating an *event*. It is also the "inactive" state (e.g., dormant volcano). The static level is the retreating "world" of events, e.g., doing something becomes a negative event of 'not doing' (a Lupascian logic term).

*2) Dynamic Model:* includes a static model subdiagram (region) that unfolds with time, leading to events, i.e., the realization of static things and actions. Thus, the event is the existing being that was formerly a subsisting being as a region in the static level. In this context, the Lupascian notion of a negative event as reverting to the static level from the dynamic level may explain Russell's [18] shadowy things, e.g., *Today is Wednesday* when in fact it is Tuesday. *Today is Wednesday* is a static (subsisting) thing that can exist (event) when *It is Tuesday*.

Before presenting the TM modeling in details, we will introduce its foundation in terms of Stoic ontology.

### A. Stoic Addition to TM Two-Level Modeling

Stoic ontology serves to define the *being* of TM things and actions in reality. The Stoics concocted the idea of a broader category of being: reality is made of things that *exist* and things that *subsist*. This idea retains the commonsensical notion that both static things and dynamic things are in some sense real. The notion of "modes of being" appears in different form in classical logic where the notions of existence and subsistence appear [19]. Reference [20] introduced Meinongian metaphysics and has distinguished between being and existence. Using Stoic ontology, we view the dynamic TM description as *in existence*, whereas the static, mapped portion of the dynamic description is *in subsistence*.

TM staticity refers to a static model that represents the world of potentialities with atemporal and nonspatial subsistence. The static model is self-contained (no new outside thimacs added) and in a state in which time and its related notions lose meaning. This *static universe* "contains everything there is or ever was or will be" (from [21] after adding "static" at the beginning).

In the dynamic TM level, events form among themselves an interacting nexus (assemblages) that define, inform, and constitute all thimac beings (existence). Things at the dynamic level may present both object-like and PROCESS-like aspects [22]. We use PROCESS with capital letters to distinguish this term from *process* action used in a TM (Fig. 2). PROCESS in a TM is another term for event and, more specifically, a net of events, as will be discussed in detail when we model BPMN.

The event can be provisionally defined as a fundamental happening that forms the basic building blocks of the existing world. Everything in the world, including people and things, can be constructed from events that form essential ontological elements. Understanding their "being" is the major field of investigation conducted by Heidegger.

### B. Example: Regions and Events

Reference [22] considered a simple homogeneous, open-ended PROCESS such as *walking*. According to [22], this seems to have a "universal" character as walking is present at many different spatiotemporal locations with many participants. But if so, what is the spatiotemporal extent of that walking? According to [22], in ordinary discourse, people typically report the occurrence of such an event as "I walked to the station," "I walked for five minutes," or "I walked 500 meters." It is not clear that we need a notion of a "process token" distinct from the widely accepted notion of an event token [22].

Fig. 3 shows the static representation of *Walking to the station*. This representation reflects a *subsisting* machine (PROCESS) that has the potential to be actualized in reality. It is subsisting because it has a being, but such a being is in the inactive state. It is not a mental thing because even if human beings vanished, it is possible that some non-human (e.g., a dog) may walk to a station. It is an inactive "natural" PROCESS like a dormant volcano that had experienced eruptions long before the existence of human beings.

Note that staticity here includes all variations of a certain thimac to permit ontological changes at the dynamic level. For example, ontological relationships among thimacs may change to a different ontological arrangement, e.g., A is no more part of B; in this case, the static level includes all these disparities.

A subsisting thimac moves to existence if it is embedded with a time subthimac to become an event. We can establish events to the generic actions (e.g., transfer, receive, etc.) in Fig. 3, but we prefer to declare more meaningful events as shown in Fig. 4. The region of an event is the subdiagram of the static model where the event occurs. For simplification, hereafter, we will use regions to indicate events.

Fig. 5 shows the event *Walked 500 meters* and Fig. 6 shows the event *Walked for five minutes*.

To illustrate how the two levels of a TM world relate to each other, Fig. 7 shows the TM PROCESS of "becoming" from a static region to an event (Event 1 is taken from Fig. 4). Accordingly, following Whitehead's metaphysics, things in reality are PROCESSes, i.e., are constituted by PROCESSes (in TM, events or net of events), and their *becoming* is also a PROCESS.

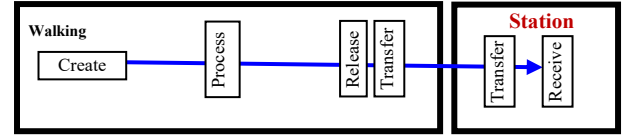

**Fig. 3 Static model of *Walking to the station***

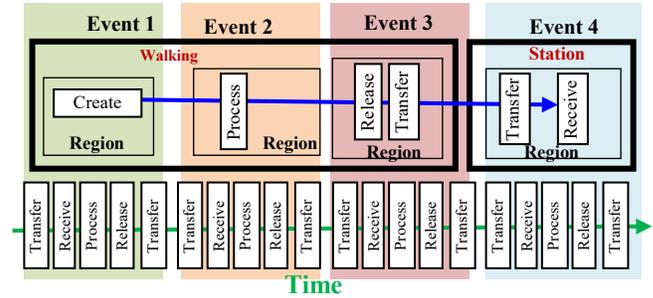

**Fig. 4 Dynamic model**

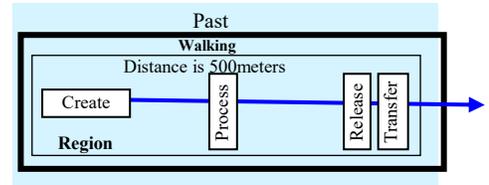

**Fig. 5 The event of *Walked 500 meters***

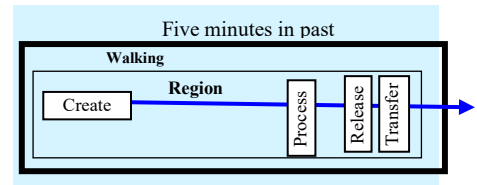

**Fig. 6 The event of *Walked for five minutes***

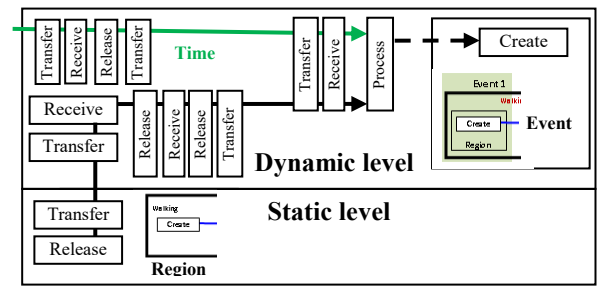

**Fig. 7 Converting a region to an event**

In the context of SCM, such issues and other related ones (e.g., nature of time, identity, and events calculus) need further investigation. Note that the development of TM and SCM is an ongoing venture further refined in each paper.

### C. The Thing Side of the Thimac

The thimac is a whole that is more than the sum of its parts (i.e., has its own machine). Even if interiority has no subthimacs (e.g., empty storage safe), the thimac has its actions: create, process, release, transfer, and/or receive. Thing subsistence means it is an identifiable thing along with its related actions. For example it is like a city in a map. The city can be described in terms of streets, population, connections with other cities, interaction with the environment, windiness, water resources, etc., but it is just a map with no activities. Even though it is connected with another city, there are no moving cars on the highways and no playing children in the streets. "Relations" between subsisting things are like dry river beds. Even though a dry river (e.g., release, transfer, transfer, receive) looks "permanent" in the static model, it becomes a flash *event* that may perish any time, i.e., alternate between static and dynamic levels.

Only thimacs that can embed time are realizable (exist) at the dynamic level. Thus, for example, "square circle" is a static thimac that cannot be injected with time to exist in the dynamic model; neither does it subsist because it is not mappable to the dynamic level. The TM universe is populated by things that alternate between two different levels of being: static and dynamic. This universe is a PROCESS (net of events) where events occur and then perish or cease to be.

According to [23], the Stoic incorporeals (mapable occupants of the static level) are conditions "without which the interaction of bodies in the world would neither be analyzable nor intelligible." The being of those incorporeals (entity-like and PROCESS-like) does not depend on their material occupants, for they can carry on in their own subsistent way without being materialized (in TM existence).

### D. The Machine

TM actions seen in Fig. 2 can be described as follows.

*1)  Arrive:* A thing moves to a machine. In Stoic ontology, motion exists along the temporal dimension and "the length of a motion, or its duration, can be measured by how much of this temporal dimension the motion covers. One motion has a longer duration than another just in case it covers more of the temporal dimension" [24].

In TM modeling, not only time is the dimension in which motion exists, but also, time is the dimension in which actions (create, process, release, transfer, and receive) exist.

*2)  Accept:* A thing enters the machine. For simplification, we assume that arriving things are *accepted*; therefore, we can combine *arrive* and *accept* stages into the *receive* stage.

*3)  Release:* A thing is ready for transfer outside the machine.

*4)  Process:* A thing is changed, handled, and examined, but no new thing results. This also includes the action of comparison, e.g., a number (thing) is greater than zero? Also consider creating a wood fire then "the wood is burning" is a

*process* indicating the physical situation has remained the same, burning.

*5)  Transfer:* A thing is input into or output from a machine. The dynamic (not necessarily physical) "movement" (event) is from a previous region to a different region through yet a third region.

*6)  Create:* A new thing "becoming" (found/manifested) is realized from the moment it arises (emergence) in a thimac. Simultaneously it also refers to the "existence" of a thing, especially where we want to emphasize persistence in time.

Ending the creation of an event returns the region of the event to the static level. There are three types of creating: existing, subsisting, and appearing (not subsisting in the static level, e.g., *square circle*). Note that "not subsisting" implies no possibility of existence (see Fig. 8). These things "neither subsist nor exist but they are still objects [things] in their own right—that is, they are part of the catalogue of the world in their own right" [25]. Creation as existence is assertive (e.g., true), creation as subsisting is expressive, creation as appearance is inexpressive in the dynamic level.

Additionally, the TM model includes a *triggering* mechanism (denoted by a dashed arrow in this article's figures), which initiates a (non-sequential) flow from one machine to another. Moreover, each action stage may have its own memory storage (denoted by cylinder in the TM diagram) of things. A memory has its own five actions forming a *memory thimac*.

Note that for simplicity, we may omit *create* in some diagrams because the box representing the thimac implies its existence (in the TM model).

### E. Two Levels of Representation and Lupascian Logic

The TM event is different from similarly named notions currently used in the literature. Note that the TM approach takes the side of philosophers (e.g., Whitehead) who conceive of physical things as extended across time. Objects and events are things of the same kind [26]. In this context, the TM differs from such an ontological approach to objects by introducing the notion of the thimac as a thing and machine extended in time.

Note that objects are elements on the dynamic level and typically viewed as durable solid things. Their static counterparts are withdrawn to the static level and physically imperceptible. Additionally, with respect to intentionality, the same object may appear partially, hence differently, at the dynamic level. However, we ignore the notion of intentionality at this stage of research.

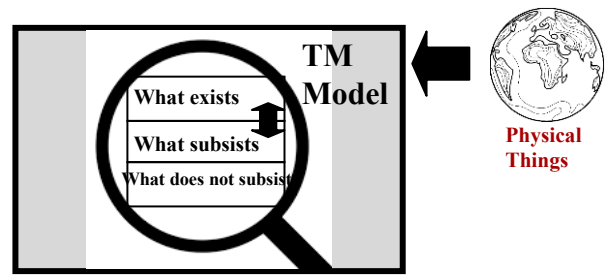

Fig. 8 Things in SCM

In SCM, instead of PROCESS vs. "stop PROCESS," the event moves the region to the dynamic level in contrast to "revert (the event's region) to static level." This method of eliminating negativity stems from philosopher Stéphane Lupasco. According to [27], every element *e* (in TM: an event, i.e., a thimac that contains a region plus time) always associates with a *non-e* (in TM: static thimac), such that the actualization of one entails the potentialization of the other and vice versa, alternatively, "without either ever disappearing completely" [27]. More illustration of this topic can be found in [15]. The next section and last section in this paper include examples of negative events represented according to Lupascian logic. Stoic ontology allows continuous passage from corporeals to incorporeals and back again as a cycle of subsistence and existence.

## III. EXPLORATION IN SCM

The Stoics' "body" (the subject matter of physics) is anything that is capable of acting or being acted upon. It is the mark of existence. A thimac is a machine that creates, processes, releases, transfers, and/or receives; simultaneously, it is a thing that can be created, processed, released, transferred, and/or received. Thus, where as in Stoicism, two principles together constitute physical reality: that which acts and that which is acted on, in TM, *acts* are creates, processes, releases, transfers, or receive; and *is acted on* are being created, processed, released, transferred, and received. In TM, thimacs can be actors and actees simultaneously.

### A. Example 1

The famous Heidegger's hammer is an existing tool as long as it is ready-to-hand. However, this existence (ready-to-hand) has two modes of being, one as a physical thing (create) and as an actually used thing (PROCESS). When it breaks down (present-at-hand), it subsists (See Fig. 9, left). What makes a hammer a "known" thing is its whole, e.g., function and utility. Note that the head and hand of the hammer exist. As Fig. 9 (left) shows, the hammer conceals itself (withdrawn) into subsistence, thus, unavailable, but its pieces exist. In Fig. 9 (middle), it exists, but it may not be used. In Fig. 9 (right), the hammer exists and is being used. Thus, it seems that present-to-hand and ready-at-hand require a third mode of being, used-at-hand. This illustrates one feature of SCM representation.

### B. Example 2

This example is more complex that previous examples in this paper. It points to a need for a calculus of events in [28], business process modeling based on TM.

According to [23], in *My watch was made in Switzerland*, the expression "my watch" refers not to a piece of metal but to an immaterial office, a role for something to be, filled by some piece of metal. If it becomes true that *Tomorrow I lose my watch and buy a new one*, then a distinct piece of metal from the one currently on my wrist will tomorrow become my watch. Therefore, *my watch* hardly refers to any piece of material.

Fig. 10 shows the TM static model for this scenario. First, a person (pink 1) buys a watch (2) to become his/her watch

(3). "Losing" is the negative event "not owning" that has the same region as the event owning but without time, as in the dynamic model. Fig. 11 shows the dynamic model that includes the following events.

$E_1$: I am an existing object (i.e., entity-like event).
$E_2$: A watch is an existing object.
$E_3$: I bought the watch.
$E_4$: The watch comes into my possession.
$E_5$: I own the watch.
R5: I do not own the watch (lose it); R denotes region 5.

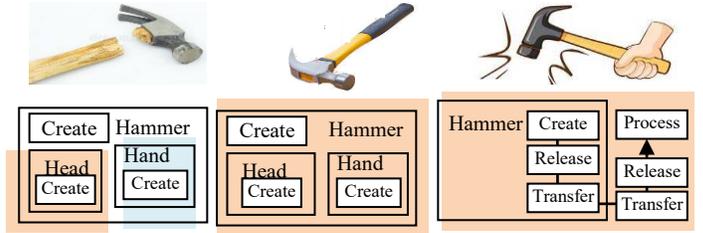

**Fig. 9 Subsisting hammer with existing head and hand (left), existing hammer (middle), existing used hammer (right)**

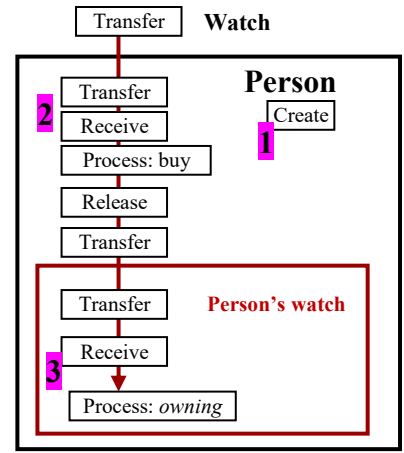

**Fig. 10 Static model of *I lose my watch and buy a new one***

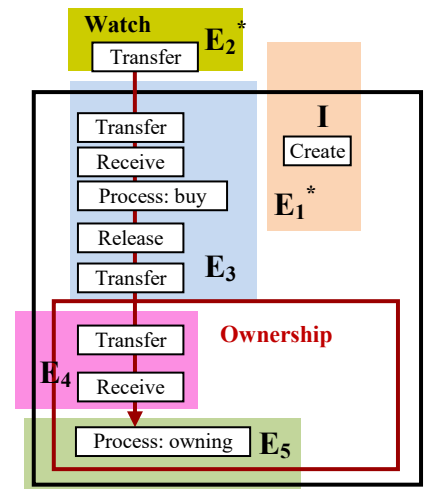

**Fig. 11 Dynamic model of *I lose my watch and buy a new one***

In Fig. 11, $E_1$ and $E_2$ indicate my and the watch's existence, and I bought the watch ($E_3$). $E_4$ denotes that the watch becomes my watch; $E_5$ registers the state of being of the owner of this watch. R5 (the region of $E_5$) indicates that I lost my watch; hence, the watch is no longer in my ownership. There is another watch in $E_2$ that I bought in $E_3$ (different time) that becomes my watch. The interesting point in this example is how the regions of $E_2$, $E_3$, and $E_4$ are repeated in different times. Hence, regions are elevated to events in different times. Asterisks indicate extended events (persisted entity-like event). It is possible to limit this extension (begin–end), but in this example, we ignore such detail.

Fig. 12 shows the behavior model.

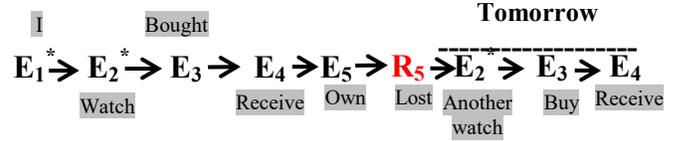

**Fig. 12 Behaviour model**

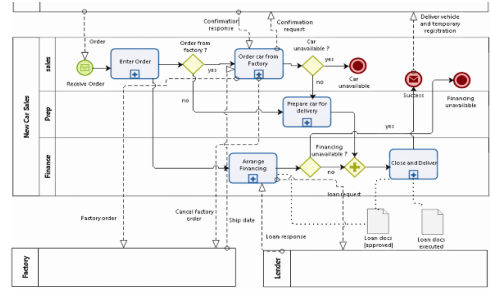

**Fig. 13 BPMN example (From [32], originally in [31])**

## IV. BPMN Using SCM

A business *PROCESS* describes how an organization performs the work necessary to produce outputs [29]. Reference [30] defines a business PROCESS as "a series of steps designed to produce a product or service." The BPMN specification defines business PROCESS as "a sequence of activities leading from an initial state of the PROCESS instance to some defined end state" [31]. An activity "is an action that is performed repeatedly in the course of business. Each instance of the activity represents the same action (more or less) on a different piece of work" [31]. Business PROCESSes may have been carefully designed, or they may have simply evolved over the years. PROCESSes are central to understanding how anything in the organization creates value [29]. Reference [29] used the famous input–process–output model to define a business PROCESS as activities that transform inputs into outputs valued by an organization.

In this paper, we develop the view that a business PROCESS is a thimac realized as a net of TM PROCESSes. The organization as a thimac can be decomposed into major PROCESSes that in turn can be divided into subPROCESSes.

### A. New Car Sale Modeling

Reference [32] presented "fact-based analysis" of BPMN example models, as shown in Fig. 13 [31]. The following concepts apply.

1) *Activity:* work performed within a business.
2) *Process:* any activity performed within or across companies.
3) *Event:* something that "happens" during the course of a business process.

Then, [32] uses such a list and Fig. 13 to develop the following fact verbalizations as ground facts (only a small subset of the verbalizations is shown here):

**The Black-Box Pool:** "Customer" has an outgoing **MessageFlow** "Order" to the **Message Start Event** "Receive Order" within the **Lane** "Sales" of **The White-Box Pool:** "New Car Sales." The **Black-Box Pool** "Customer" has an outgoing **MessageFlow** "confirmation response" to the **SubProcess** "Order Car from Factory." The **SubProcess** "order Car from Factory" has **An outgoing MessageFlow** "confirmation request" to the **Black-Box Pool** "Customer."

Such an approach is claimed to enhance the "semantic richness" of BPMN. Reference [32]'s example provides a typical example of a business PROCESS, meaning a network of TM events. We claim that such an attempt to provide more BPMN semantics is caused by the modeling in BPMN. To demonstrate that, we develop a model based on SCM in order to contrast the two models side by side.

### B. Static Model

Fig. 14 shows the TM static model (originally in the subsisting state) of regions of the new car sale PROCESS that can be described as follows.

A customer creates an order for a new car (circle 1) that flows to the new car sale (2) where it is processed (3). According to the result of such a process,

a) If the car is not available (4), then this triggers creating a message (5) that flows to the customer (6).

b) If the car is to be ordered from the factory (7), then a request to confirm is sent to the customer (8), and with getting this confirmation (9, 10 and 11), a request is sent to the manufacturer (12) and a request is sent to the preparation section (13) to prepare the car.

c) If the car is available (14), a request is sent to the preparation section (15).

In both cases of (b) and (c) above, a request for finance is sent to the finance section (16 and 17). In the finance section, the request for financing is processed (18) to trigger generation of a loan request (19) that flows to a loaner (20).

The loaner processes the loan request (21) and sends the response to the finance section (22). The finance section processes the loaner response (23) and sends an approval/rejection decision to the preparation section (24). In the preparation section,

a) If a preparation request (25) and a finance approval (26) are received, the car is taken out of storage (27), prepared (28), and sent to the customer (29). The horizontal thick bar is a graphical simplification that indicates satisfying both conditions in 25 and 26.

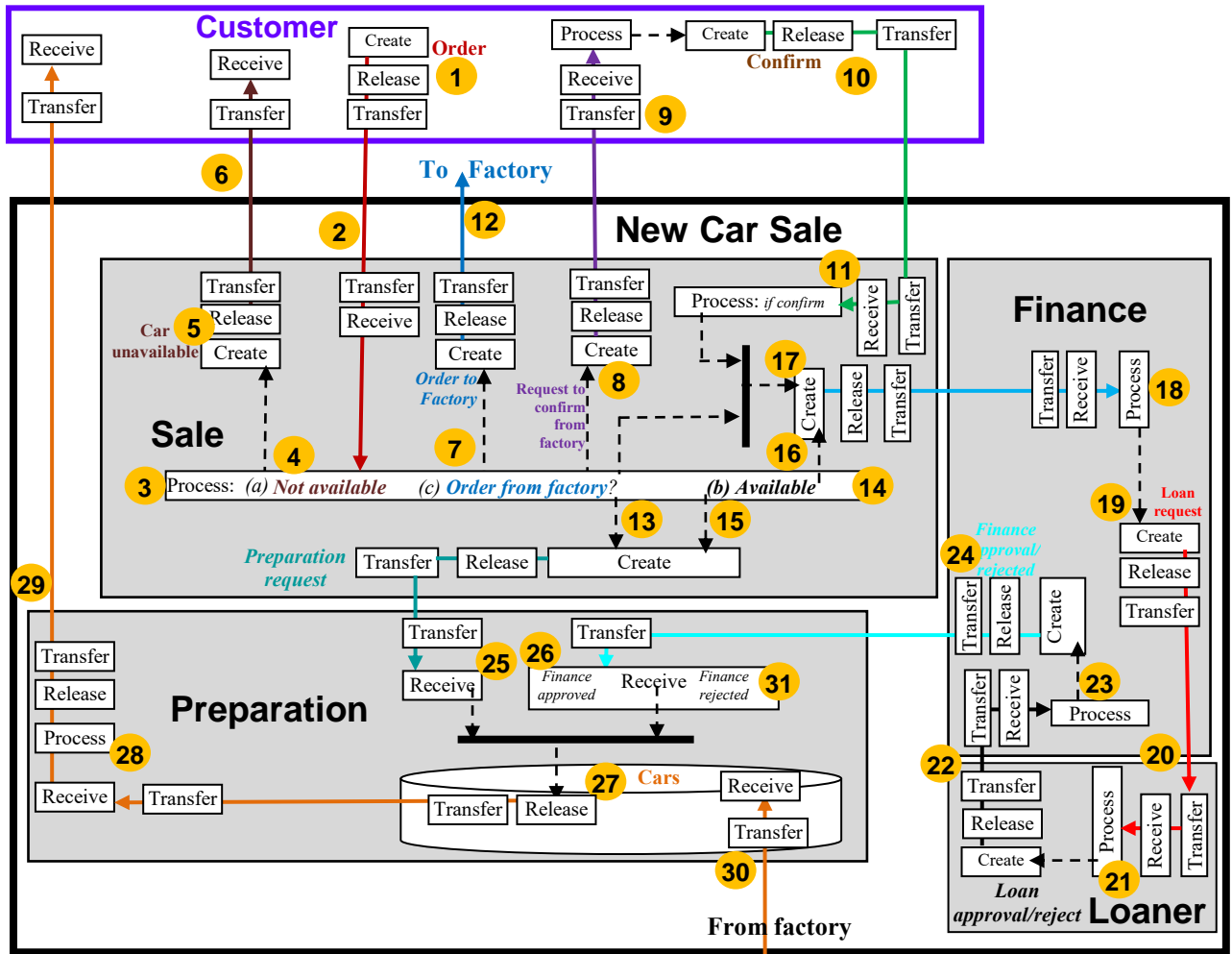

**Fig. 14 The static model of the new car sale**

In the case of a car ordered from the manufacturer, we assume a third condition that the car has arrived to the storage from the manufacturer (30).

b) In (a) above, if the finance is rejected (31), then the actions in (a) are not executed (negative event as will be indicated in the dynamic model).

### C. Dynamic Modeling

Fig. 15 shows the dynamic model with the following events.

$E_1$: The customer sends an order for a new car that is received by the new car sale.

$E_2$: Processing the customer's order.

$E_3$: The ordered car is unavailable; hence, a message is sent to the customer informing him/her.

$E_4$: Ordering a car involves a car from the factory.

$E_5$: Making order to factory.

$E_6$: Requesting and receiving confirmation of ordering from factory.

$E_7$: Sending a request for preparation of a car.

$E_8$: Sending a request for financing the car.

$E_9$: The car is available.

$E_{10}$: Creating a request for a loan that flows to a loaner.

$E_{11}$: The loaner responds to the loan request.

$E_{12}$: The finance section processes the loaner response and sends approval/rejection to the preparation section.

$E_{13}$: The preparation section receives an approval of financing.

$E_{14}$: The preparation section receives a preparation request.

$E_{15}$: The car has arrived from the factory.

$E_{16}$: The car is delivered to the customer.

$E_{17}$: Finance is rejected.

$R_{16}$: (negative event) The car delivery is cancelled.

Fig. 16 shows the chronology of these events.

Thus, the new car sale PROCESS starts at the subsistence (static) level where no events occur. Note the sense of subsistence in this initial step. All regions have beings (they are real), but they are inactive. For example, if a customer asks, do you sell used cars? The answer is *there is* no such PROCESS in our company.

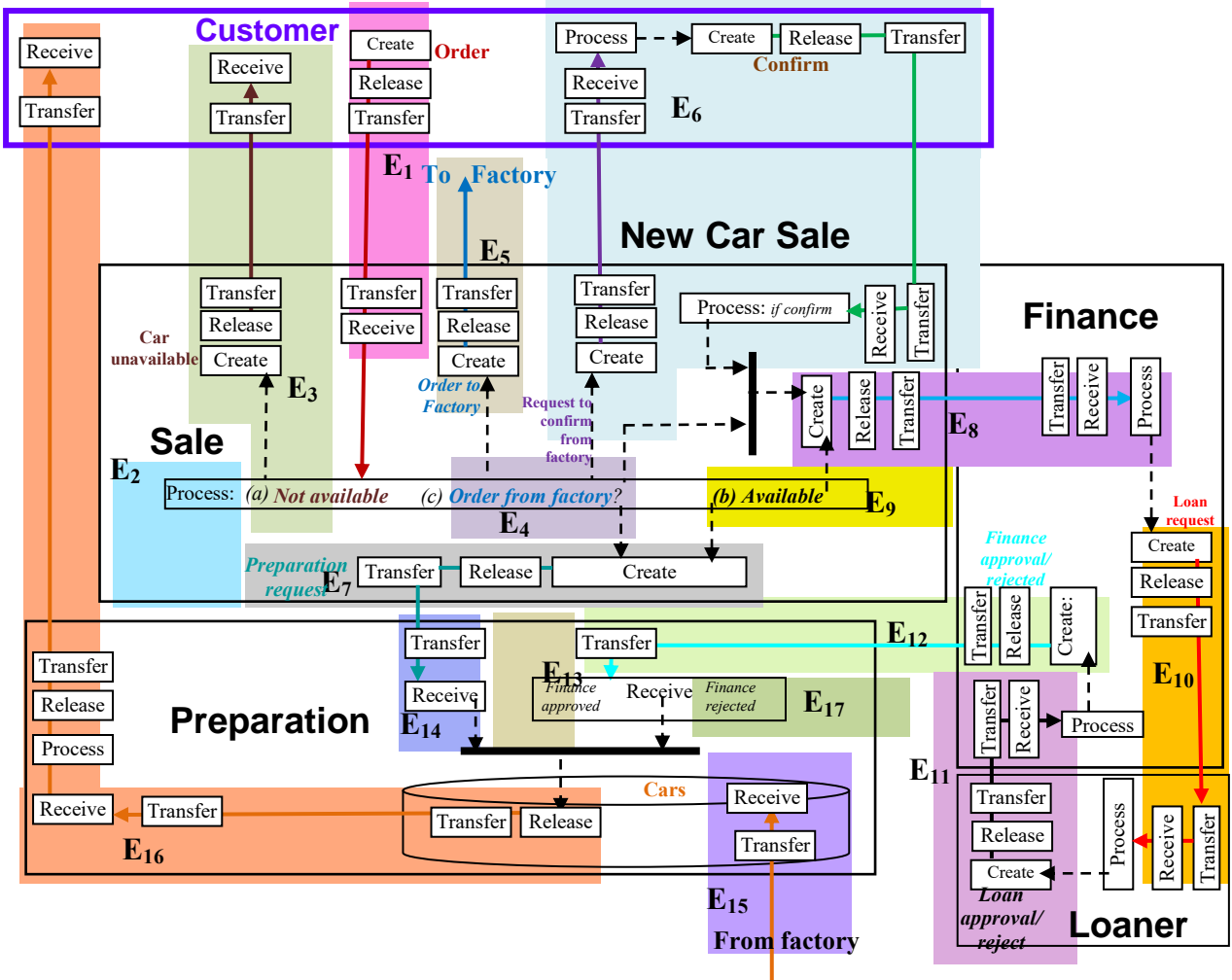

**Fig. 15 The dynamic model of the new car sale**

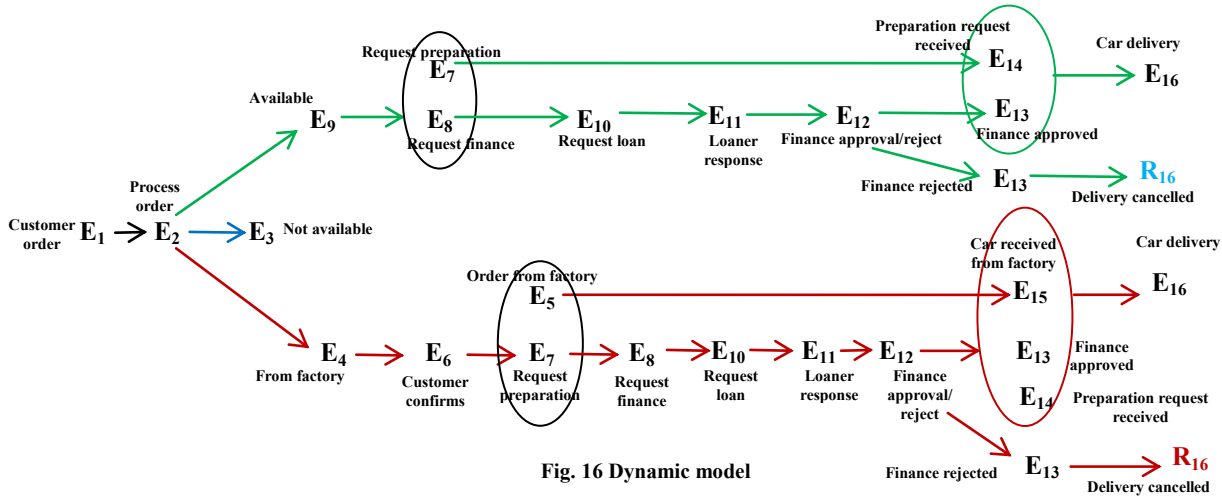

**Fig. 16 Dynamic model**

Note how the events arise from the static level in sequences for a single order, e.g., ordering available car (see Fig. 17). Of course, there are several simultaneous orders.

All types of studies can involve the number of events that are currently occurring per region, the number of regions (negative events), etc. that provide data for efficiency and activities across all PROCESSes.

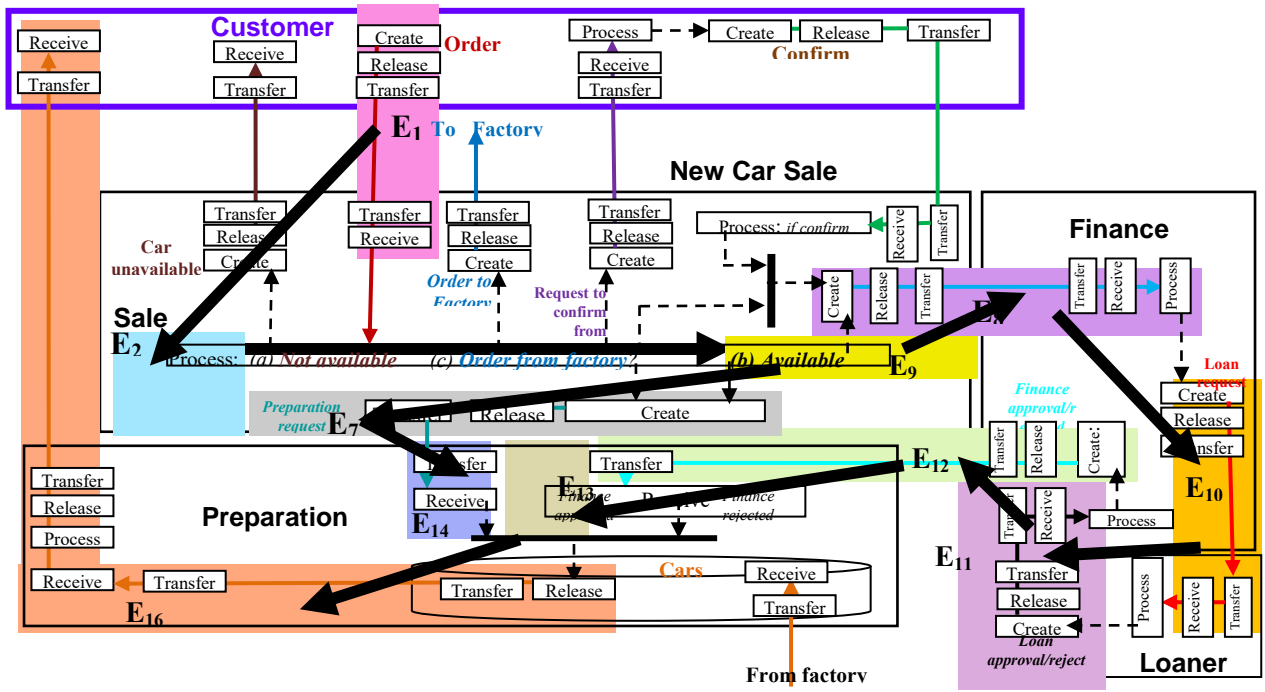

Fig. 17 The trace of events for available car

## V. CONCLUSION

As a case study, we have applied SCM to investigate the ontology of BPMN. Such an undertaking would demonstrate the SCM notions and simultaneously may offer a viable ontological foundation for BPMN. Specifically pinpointing the disadvantages and disadvantages between the two modeling techniques is difficult because they differ in every aspect. The strategy is to compare them by modeling the same problem in the two approaches in parallel and examine the overall results.

The judgement is left for the reader in such matters, for example, in [10], the ability "to describe and reflect the real world of information systems better." Further research is needed to clarify the feasibility of SCM-based modeling.

However, one issue may be raised about the diagrammatic "complexity" of the TM model. Such a complexity seems to be caused by the repeated use of the TM actions. For example, by assuming that the direction of arrows is sufficient to indicate the flow, Fig. 14 can be simplified as shown in Fig. 18 that eliminates the actions release, transfer, and receive.

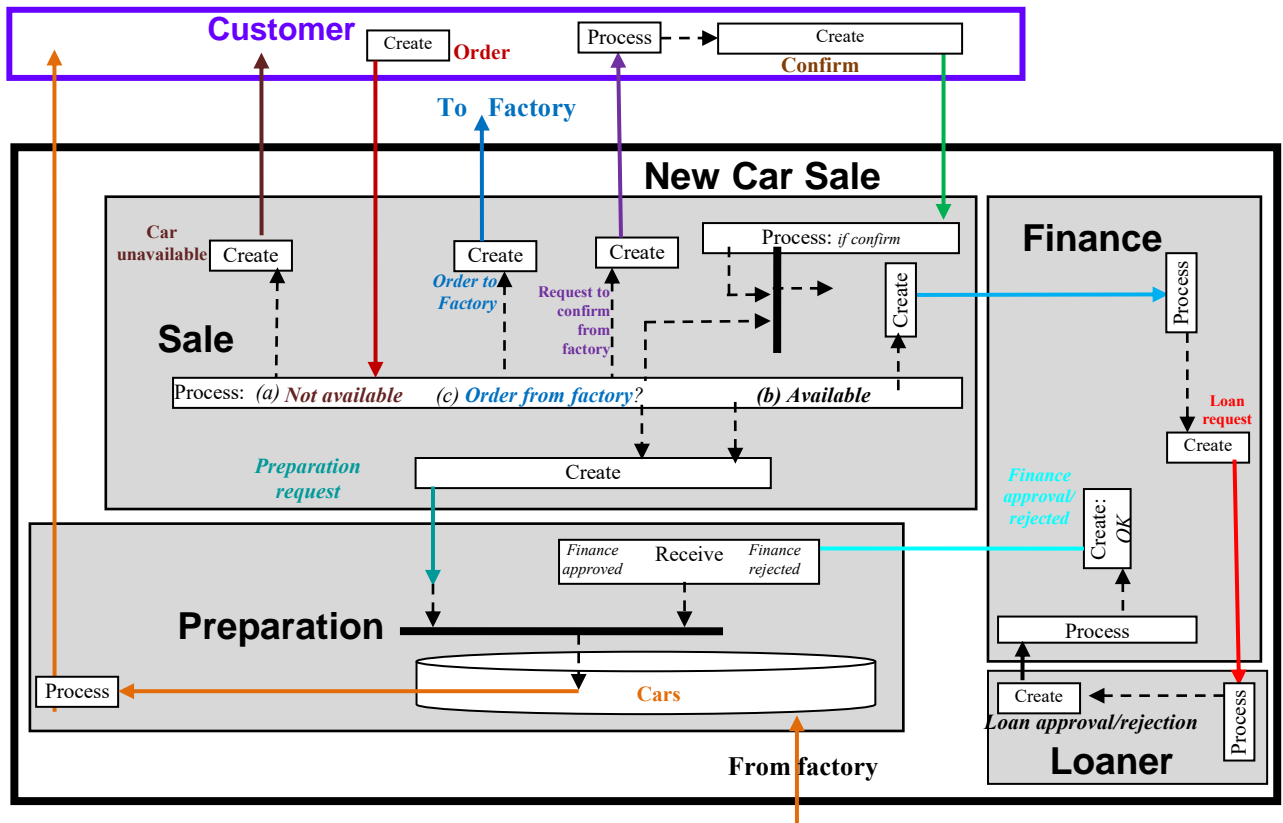

**Fig. 18 Simplification of the static model of the new car sale**